\documentclass[sigconf,nonacm]{acmart} %
\usepackage{amsmath,amssymb,amsfonts,chemarrow}
\usepackage{hyperref}
\usepackage{graphicx}
\usepackage{subfigure}
\usepackage{url}
\usepackage{mathenv}
\usepackage{color}
\usepackage{breqn}
\usepackage{fancybox}
\usepackage{algpseudocode}
\usepackage[linesnumbered,ruled]{algorithm2e}
\newcommand{\BfPara}[1]{\vspace{1mm}{\noindent\bf#1.}\xspace}
\usepackage{multirow}
\usepackage{float}
\usepackage{threeparttable}
\usepackage{longtable}
\usepackage{supertabular,booktabs}
\usepackage{algpseudocode}
\usepackage{booktabs}
\usepackage{graphics}
\usepackage{epsfig}
\usepackage{listings}
\usepackage{rotating}
\usepackage{amsthm}
\usepackage{hyperref}
\usepackage{xcolor}
\usepackage{upquote}
\usepackage{xspace}
\usepackage{setspace}
\usepackage{caption}
\usepackage[inline]{enumitem}
\usepackage{mathenv}
\usepackage[english]{babel}
\newcommand{\vs}[1]{{\vspace{-#1mm}}}
\usepackage{balance}
\usepackage{natbib} 
\usepackage{framed}
\usepackage[T1]{fontenc}
\usepackage[utf8]{inputenc}
\usepackage{environ}
\usepackage{tikz}
\usetikzlibrary{arrows}
\usetikzlibrary{calc,matrix}
{\bfseries}{\itshape}
{\bfseries}{\itshape}
{\bfseries}{\itshape}
{\bfseries}{\itshape}
{\bfseries}{\itshape}
{\itshape}
\setcopyright{none}
\usepackage{pifont}
\newcommand{\ours}{{COPYCAT}\xspace}
\usepackage{xspace}
\usepackage{setspace}
\newcommand{\note}[1]{}

\newcommand{\etc}{{etc.}\xspace}
\newcommand{\eg}{{\em e.g.}\xspace}
\newcommand{\ie}{{\em i.e.,}\xspace}
\DeclareMathOperator*{\argmax}{argmax}

\newcommand{\etal}{{\em et al.}\xspace}

\usepackage{hyperref}
\definecolor{linkcolour}{rgb}{0,0.2,0.6}
\definecolor{xgreen}{rgb}{0.2,0.6,0.0}
\definecolor{xred}{rgb}{0.7,0.1,0.0}
\def\equationautorefname~#1\null{(#1)\null}

\usepackage{tikz}
\usetikzlibrary{shapes.geometric, arrows}
\tikzstyle{startstop} = [rectangle, rounded corners, minimum width=3cm, minimum height=1cm,text centered, draw=black, fill=red!30]
\tikzstyle{io} = [trapezium, trapezium left angle=70, trapezium right angle=110, minimum width=3cm, minimum height=1cm, text centered, draw=black, fill=blue!30]
\tikzstyle{process} = [rectangle, minimum width=3cm, minimum height=1cm, text centered, draw=black, fill=orange!30]
\tikzstyle{decision} = [diamond, minimum width=3cm, minimum height=1cm, text centered, draw=black, fill=green!30]
\tikzstyle{arrow} = [thick,->,>=stealth]

\colorlet{punct}{red!60!black}
\definecolor{background}{HTML}{ffffff }
\definecolor{delim}{RGB}{20,105,176}
\colorlet{numb}{magenta!60!black}
\definecolor{light-gray}{gray}{0.95}
\definecolor{darkgray}{rgb}{0.4, 0.4, 0.4}
\definecolor{editorGray}{rgb}{0.95, 0.95, 0.95}
\definecolor{editorOcher}{rgb}{1, 0.5, 0} 
\definecolor{editorGreen}{rgb}{0, 0.5, 0} 
\definecolor{orange}{rgb}{1,0.45,0.13}      
\definecolor{olive}{rgb}{0.17,0.59,0.20}
\definecolor{brown}{rgb}{0.69,0.31,0.31}
\definecolor{purple}{rgb}{0.38,0.18,0.81}
\definecolor{lightblue}{rgb}{0.1,0.57,0.7}
\definecolor{lightred}{rgb}{1,0.4,0.5}
\usepackage{upquote}
\usepackage{listings}
\definecolor{pblue}{rgb}{0.13,0.13,1}
\definecolor{pgreen}{rgb}{0,0.5,0}
\definecolor{pred}{rgb}{0.9,0,0}
\definecolor{pgrey}{rgb}{0.46,0.45,0.48}

\hypersetup{colorlinks,breaklinks,citecolor=blue!50!black, urlcolor=red!50!black, linkcolor=green!50!black}

\begin{document}

\title[\ours: Practical Adversarial Attacks on Visualization-Based Malware Detection]{\ours: Practical Adversarial  Attacks \\ on  Visualization-Based Malware Detection}

\author{Aminollah Khormali}
       \affiliation{\institution{University of Central Florida}}
       \email{aminkhormali@Knights.ucf.edu}
       
\author{Ahmed Abusnaina}
       \affiliation{\institution{University of Central Florida}}
       \email{ahmed.abusnaina@knights.ucf.edu}

\author{Songqing Chen}
       \affiliation{\institution{George Mason University}}
       \email{sqchen@gmu.edu}

\author{DaeHun Nyang}
       \affiliation{\institution{INHA University}}
       \email{nyang@inha.ac.kr}
       
\author{Aziz Mohaisen}
       \affiliation{\institution{University of Central Florida}}
       \email{mohaisen@ucf.edu}

\begin{abstract}
Despite many attempts, the state-of-the-art of adversarial machine learning on malware detection systems generally yield unexecutable samples. In this work, we set out to examine the robustness of visualization-based malware detection system against adversarial examples (AEs) that not only are able to fool the model, but also maintain the executability of the original input. As such, we first investigate the application of existing off-the-shelf adversarial attack approaches on malware detection systems through which we found that those approaches do not necessarily maintain the functionality of the original inputs. Therefore, we proposed an approach to generate adversarial examples, \ours, which is specifically designed for malware detection systems considering two main goals; achieving a high misclassification rate and maintaining the executability and functionality of the original input. We designed two main configurations for \ours, namely AE padding and sample injection. While the first configuration results in untargeted misclassification attacks, the sample injection configuration is able to force the model to generate a targeted output, which is highly desirable in the malware attribution setting. We evaluate the performance of \ours through an extensive set of experiments on two malware datasets, and report that we were able to generate adversarial samples that are misclassified at a rate of 98.9\% and 96.5\% with Windows and IoT binary datasets, respectively, outperforming the misclassification rates in the literature. Most importantly, we report that those AEs were executable unlike AEs generated by off-the-shelf approaches. Our transferability study demonstrates that the generated AEs through our proposed method can be generalized to other models. 

\end{abstract}

\keywords{Malware Detection; Visualization; Adversarial Examples; Deep Learning}

\maketitle 

\section{Introduction}
As the threat of malicious software (malware) is growing rapidly, malware detection is becoming a higher priority within the computer security research community and software security industry alike.  To address this threat, there has been a wide variety of research works presenting various techniques for detection, using signatures~\cite{FarukiLBGG15}, static analysis~\cite{kang2015detecting}, and dynamic analysis~\cite{bulazel2017survey}. Although these methods are powerful in analyzing malware, they have multiple shortcomings, such as their limited scalability~\cite{Al-DujailiHHO18}. The static analysis based approach to malware detection is ineffective, since it is prone to obfuscation. Dynamic analysis, while addressing obfuscation by observing actual behavior of malware at the execution time, is often resources-intensive, and subject to dynamic analysis evasion (e.g., sandbox detection), making it insufficient alone. Those shortcomings and the need for an automation technique that extrapolates insight into software from features of behavior have opened a new avenue for malware analysis through visualization, often incorporated into automatic malware detection techniques that often use machine and deep learning~\cite{natarajKSGM2011,HanLKI15,CappersMEW18,AngeliniBBCDFLS18}. 

Visualization has been leveraged over both shallow and deep learning algorithms, and in many security applications, including computer forensics~\cite{ChenCAANTTY18}, network monitoring~\cite{GomezNA17}, and software security~\cite{HanLKI15,CappersMEW18,AngeliniBBCDFLS18}. In malware analysis of visualization techniques, hard-to-analyze code information is transformed into an image, where features of different executable files are visualized. Compared to conventional methods, including those mentioned above, the detection process using visualization is simpler and faster (analogous to static analysis), due to the viability of deep learning approaches in recognizing patterns, even over obfuscated code~\cite{AbuhamadAMN18}. Moreover, the utilization of visualization techniques along with deep learning models has been shown to improve the performance of malware detection in multiple ways: reduced time and memory consumption, improved robustness and detection accuracy, and being non-intrusive~\cite{HanLKI15, makandar2017malware, GrossePMBM17, SuVPSFS18}. To this end, several studies have incorporated deep learning-based networks for detecting malicious executable files from benign ones using visualization techniques~\cite{CuiXCCWC18, FuXWLS18}. 

Although applications of deep learning models are actively explored in a wide range of applications, \eg, health-care, industry, and cyber-security~\cite{AlipanahiDWF15, MohaisenAM15}, deep learning itself has been shown to be vulnerable to Adversarial Examples (AEs)~\cite{Moosavi-Dezfooli16,AbusnainaKAPAM}. AEs, which are carefully crafted inputs, are created by adding a small perturbation to the original input of a machine learning algorithm in order to produce the adversaries' desirable outputs, such as misclassification~\cite{PapernotM0JS16,KhormaliANYM19}. Since the crafted samples are generated by applying limited changes on the original inputs, the crafted samples are very similar to the original ones, and are not necessarily outside of the training data manifold. Algorithms crafting adversarial samples are designed to minimize the perturbation, thus making it hard to distinguish adversarial samples from legitimate ones. 

\BfPara{Shortcomings of the Prior Work} There has been several studies, some of which are concurrent to this work, which have several shortcomings which we outline here and address in the subsequent sections. Most importantly, while generating adversarial samples, it is important to ensure executability of the resulting samples (code), {\em a common shortcoming of the prior work}. Liu \etal \cite{liu2019atmpa} presented adversarial attacks on machine learning-based malware visualization detection methods, although they considered the entire sample as a candidate for perturbation, corrupting the adversarial sample's code. Of note in this direction of generating adversarial samples while not corrupting the sample is the seminal work of Grosse \etal \cite{GrossePMBM17}, who investigated adversarial attacks on neural network based android malware detection models. In their work, perturbations were added into malware's none functional parts. Their approach achieved a misclassification rate of 63\% over all malware samples, which is---although promising---quite low. Moreover, although their method maintains the functionality of the code, it does not necessarily maintain its executability, since changing the application  manifest will affect  the  permissions, software and hardware components access, intents, \etc, which are necessary for the malicious applications to execute the code by the correct API calls. 

Although adversarial learning has been an active research area, and outside of the aforementioned works, there is very little research done on understanding the impact of adversarial attacks on deep learning-based malware detection~\cite{GrossePMBM17, KolosnjajiDBMGE18, kreuk2018deceiving}, particularly those utilizing visualization. In parallel to our work, the authors of \cite{KolosnjajiDBMGE18, kreuk2018deceiving} suggested injection of a sequence of bytes for AE generation. Our work differs in multiple points. First, our work tackles the visualization-based malware detection approach, which has been proven to be faster and more efficient as a malware detection method~\cite{GrossePMBM17, SuVPSFS18}. Second, we investigate the targeted misclassification attacks, in addition to untargeted misclassification which was the focus of the work in~\cite{KolosnjajiDBMGE18, kreuk2018deceiving}. Third, we explore and confirm the generalization of the proposed method by evaluating it using two malware datasets,  Windows and IoT binaries, and transferability to other algorithms. Finally, we explore the performance of a defense method, adversarial training, on the detection of the generated adversarial malware examples. 

\BfPara{Contributions}
Motivated by the aforementioned issues, our main goal is generating {\em adversarial malware samples that (1) fool the classifier and (2) execute as intended.} First, we investigate the executability of the adversarial examples generated by leveraging adversarial attack methods into deep learning-based malware visualization detection systems. In particular, we explore multiple well-established adversarial attack methods with the visualization-based approach to generate adversarial samples with least $L_2$-distance from original input for both attacks and defenses. We find that although the generated adversarial examples can successfully fool the classifier, none of them is still executable.
Therefore, we further propose \ours, which operates by padding binaries of a certain class into the binaries of another label, to maintain the executability of the original input, while forcing the model to act as desired by the adversary, with targeted and untargeted misclassification. Specifically, the contributions are:

\begin{itemize}[leftmargin=*]
    \item  We investigate the application of five generic adversarial methods for attacking visualization-based deep learning models for the malware detection. We find that although these methods outperform state-of-the-art in their evasion rate, the generated AEs do not necessarily maintain the executability of the files. We conclude that those adversarial learning algorithms are impractical candidates for understanding the robustness of such detectors.

    \item  Given the shortcomings of those off-the-shelf adversarial  algorithms for generating practical malware samples that evade detectors, we propose a new approach, called \ours, which is specifically designed for generating executable samples, in order to maximize the attack success rate while maintaining the functionality and executability of the original files.

    \item We conduct comprehensive experiments to evaluate the performance of \ours for generating adversarial executable samples under both targeted and untargeted attacks settings. In particular, we have designed and evaluated various configurations of \ours using two malware datasets, for Windows and IoT, demonstrating the effectiveness of the proposed approach in generating executable AEs. \ours is shown to have a misclassification rate of 98.9\% and 96.5\% on Windows and  IoT  malware  datasets, respectively, and all of the AEs are executable.
\end{itemize}

\BfPara{Organization}
In~\autoref{sec:Preliminary} we provide a brief background. We describe \ours in~\autoref{sec:methedology}, and follow it with performance evaluation and discussion  in~\autoref{sec:Results}. We review the related work in \autoref{sec:related}. Finally, limitations and future work are discussed in~\autoref{sec:Limitation}, followed by the conclusion in~\autoref{sec:Conclusion}. 

\section{Background and Preliminaries}\label{sec:Preliminary}
We incorporate malware visualization from the computer-vision community, and the deep learning algorithms and adversarial attacks from the artificial intelligence community into a single process for generating adversarial malware samples. In this section, we describe some background information and basic models. First, we need to transform malware binaries into images from which features are extracted, as described in~\autoref{sec:prelim_visualization}. In~\autoref{sec:prelim_AMLattack}, we describe the high-level idea of the adversarial machine learning and its importance in the domain of deep learning-based malware detection systems. We note that we will provide more technical details of AEs generation methods in~\autoref{sec:method_attacks}. Finally, we discuss potential defenses against AEs in~\autoref{sec:prelim_AMLtrain}.

\subsection{Deep learning-based Malware Detection} \label{sec:metheod_DLmodel}

In~\ours, and as a baseline, we design a deep learning-based classifier based on the Convolutional Neural Network (CNN) architecture. CNNs are a type of deep, feed-forward artificial neural networks, widely used for image classification tasks~\cite{krizhevsky2012imagenet, CiresanMS12}. Deep learning models offer several advantages compared to traditional machine learning algorithms. For example, they extract and learn representative features automatically, which in turn means  that they require minimal pre-processing effort compared to traditional machine learning methods where features have to be extracted through feature-engineering algorithms using domain knowledge~\cite{DLvsML17}. 

We designed two CNN models, one for each dataset, with three consecutive convolutional layers with ReLu activation, followed by a fully connected layer outputting softmax function values for each class. We tested different values for the number of convolutional layers, batch size, and epochs to improve the classification accuracy of the model over test samples of the two malware datasets. For the Windows-based malware binaries we achieved 99.12\% accuracy rate after 50-epoch training and batch size of 150. Similarly, we trained another CNN model with the same architecture for IoT malware binaries and achieved 99.67\% accuracy rate on test samples after 50-epoch training and a batch size of 150.
More detailed information about convolutional neural networks can be found in~\cite{krizhevsky2012imagenet}.

\subsection{Visualization} \label{sec:prelim_visualization}

The different sections of code containing specific information about the software (malware) can be converted and organized as a two-dimensional array of unsigned integers, which is then transformed into a grayscale image. Recently, various visualization methods~\cite{HanLKI15, ZhangQYOH16, makandar2017malware} have been proposed to analyze malware using this idea. The main goal of malware visualization techniques is conversion of hard-to-identify code information into an image, and acceleration of malware detection process~\cite{lee2011study}. The visualization process starts reading the malware samples as 8-bit unsigned integer vector, and categorizing it into set of two-dimensional array. The values in the array are transformed into a range of $0$ and $255$ that represents a grayscale image. The pixels corresponding to $0$ result in black pixels while $255$ result in white pixels. It is worth noting that the images' width is determined and fixed based on the characteristics of the dataset, while the height of the image can change based on the size of the malware sample. The output of the visualization step would be images with the same width and probably different heights. However, before applying the learning algorithm, we need re-scale the images to a fixed width and height.

\begin{figure}[t]
\centering
\includegraphics[width=0.45\textwidth]{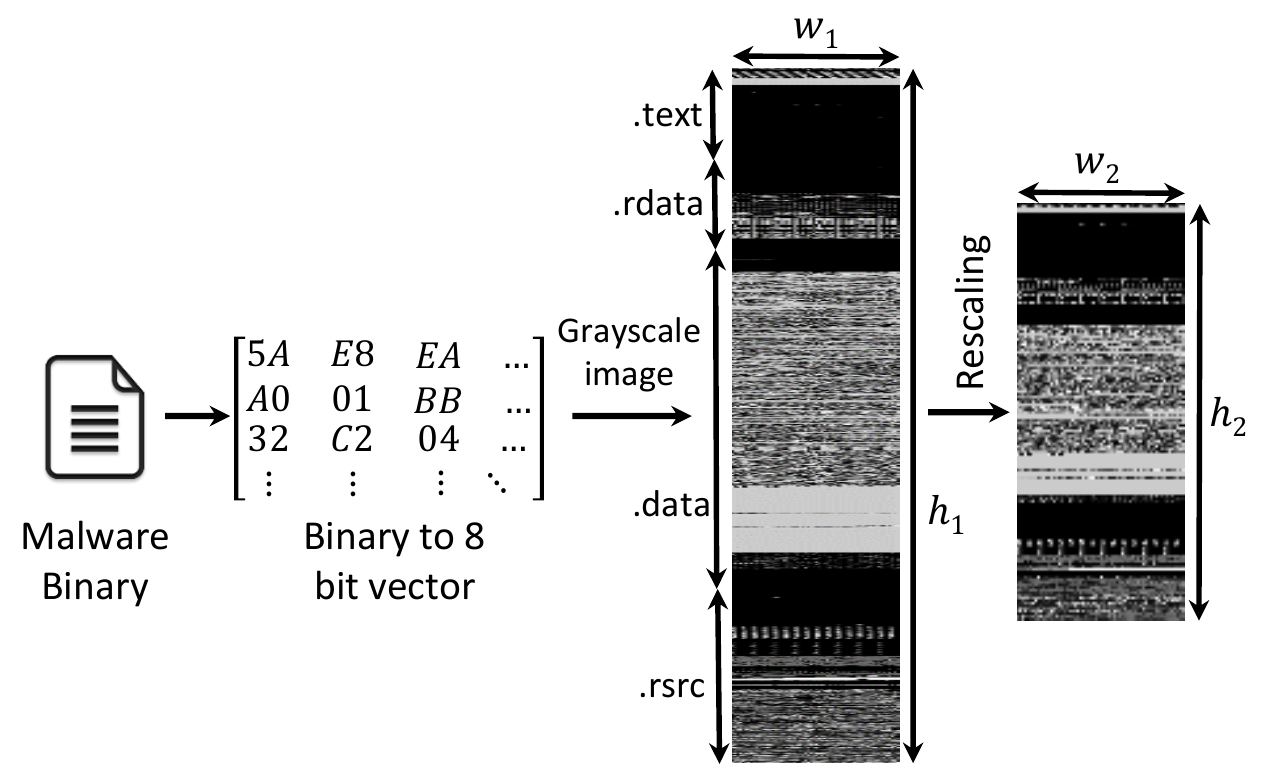}
\caption{High-level flowchart of the visualization process. Note $w_1=w_2$, while not necessarily the case for $h_1$ and $h_2$.}
\label{fig:Mal_Visualization}
\end{figure}

\BfPara{Illustration} A high-level flowchart of the malware visualization process is illustrated in~\autoref{fig:Mal_Visualization}. In this process, each malware binary is first converted into an 8-bit vector and then transformed into a grayscale image with a fixed width of $w_1$ (vector to matrix conversion). The height of the image depends on the original malware size. However, the re-scaled images of all binaries are all of the same width and height, where $w_1$ and $w_2$ are equal while $h_1$ and $h_2$ are not necessarily equal. 

\begin{figure}[t]
\centering
		\subfigure[Win. ben.\label{fig:W_benign}] {\includegraphics[width=0.09\textwidth]{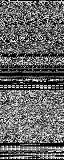}}\hspace{3mm}
		\subfigure[Win. mal. \label{fig:W_malicious}] {\includegraphics[width=0.09\textwidth]{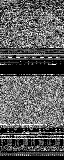}}\hspace{3mm}
		\subfigure[IoT ben. \label{fig:IoT_benign}] {\includegraphics[width=0.09\textwidth]{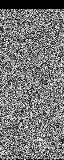}}\hspace{3mm}
		\subfigure[IoT mal. \label{fig:IoT_malicious}] {\includegraphics[width=0.09\textwidth]{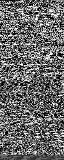}}
\caption{Sample visualization of benign and malicious binaries. Here,~\autoref{fig:W_benign} and \autoref{fig:W_malicious} show benign and malicious Windows binaries, respectively. While, benign and malicious IoT binaries are shown in~\autoref{fig:IoT_benign} and~\autoref{fig:IoT_malicious}, respectively.}\label{fig:Sample_Visualization}
\end{figure}

We note that each binary file consists of various sections, such as {\tt .text}, {\tt .rdata}, {\tt .data}, {\tt .rsrc}~\cite{conti2010visual}. The {\tt .text} section contains the executable code, the {\tt .data} section includes both uninitialized code and initialized data, while the {\tt .rsrc} section contains all resources of the module. All sections are combined to construct the image corresponding to the binary, and four different samples from our benign and malware samples are visually illustrated in~\autoref{fig:Sample_Visualization}, where \autoref{fig:W_benign} and~\autoref{fig:W_malicious} show benign and malicious Windows binaries, while ~\autoref{fig:IoT_benign} and~\autoref{fig:IoT_malicious} show benign and malicious IoT binaries, respectively. As seen in~\autoref{fig:Sample_Visualization} different fragments of benign and malware samples represent unique image textures, which are leveraged in the deep learning model to detect malicious binaries~\cite{ronen2018microsoft}. 

\subsection{Attacks in Deep Learning (Threat Model)} \label{sec:prelim_AMLattack}

The main goal of the attacks against deep learning is to modify an input sample $x$ such that it is misclassified by the model $f$, where it has been shown that an attacker is able to craft AEs, e.g., $x'$, by applying a small perturbation to the original input such that the model is forced to produce an output of the attacker's choice~\cite{Carlini017, wang2018great, PapernotMGJCS17}. Adversarial learning attacks on deep learning models can be categorized from multiple point of view, and based on the adversary's goals and capabilities~\cite{PapernotMJFCS16}, as highlighted in the following.

\subsubsection{Goals} The goal of the adversary in deep learning is to force the model into incorrect results. The main goal of the adversary can be represented based on the nature of the incorrectness. 

\begin{itemize}[leftmargin=*]

\item {\bf Confidence reduction.} The goal is to reduce the classifier's confidence, by presenting ambiguity in the predictions of the model. This attack does not necessarily result in incorrect outputs. 

\item {\bf Untargeted misclassification.} Producing adversarial examples, e.g., $x'$, that force the model output $f(.)$ to be any class other than the original one; this is, $f(x')\neq f(x)$. 

\item {\bf Targeted misclassification.} Crafting adversarial examples, e.g., $x'$, that force $f(.)$ to be any specific class $f(x')= t$. Note that this type of attack is more complex than the confidence reduction and the untargeted misclassification attacks. 
\end{itemize} 

\subsubsection{Capabilities}
Considering attacks conducted at the test time, the adversarial attacks can be categorized based on the information and capabilities made available to the adversary into white-box attacks and black-box attacks.

\begin{itemize}[leftmargin=*]
\item \BfPara{Model architecture and training data} The adversary has full knowledge and access to the training data $x$ and the deep learning model $f$ used for classification, including the training functions and algorithms, the number and type of layers, neurons' activation functions, link weights and bias, loss, \etc
\item \textbf{Model architecture} In this work we assume that the adversary has full knowledge about the model architecture and associated activation functions, the number of layers, link weights, bias, \etc 
\item \textbf{Training data} In this work we assume that the adversary builds a new deep learning model based on his collected dataset that is representative of the original dataset in both characteristics and distribution. Note that the adversary has no information regarding the original model and its properties.
\item \textbf{Oracle} The adversary has an oracle access to the model and no prior knowledge about the architecture and its internals. Moreover, we assume that the adversary can learn about the relationship between supplied inputs and the corresponding output of the model through rate limited/unlimited queries. 
\item \textbf{Sample} Here the adversary is able to collect pairs of inputs and outputs associated with an unknown deep learning model. However, the adversary cannot change the input value to investigate its impact on the model output. Therefore, it is only practical when large number of samples are available. 
\end{itemize}
\subsubsection{Our Threat Model} In this work we designed two sets of configurations for \ours depending on the utilized approach. Namely, while the AE padding approach assumes that the adversary has full knowledge of the topology of the model, the link weights, \etc, the sample injection configuration is a black-box attack. Moreover, we assume that the adversary's goal is to conduct both targeted and untargeted attacks. Based on that, part of this study falls into the standard white-box attack model, while the other part falls into black-box attack model. 

\subsection{Robustness} \label{sec:prelim_AMLtrain}
Both shallow and deep learning models play a critical role in detecting malicious applications and analyzing sequential streams of data. Therefore, several defenses have been proposed to increase the robustness of such models against adversarial learning attacks~\cite{tsipras2018there}. The ultimate goal for defensive techniques is making the decision based on meaningful features, and more immune to changes in features that are less important~\cite{tsipras2018there}. Adversarial training is the most common defensive technique, and is implemented by training the model over the original data and the generated adversarial examples from different attack methods. This defense technique ensures better accuracy against such attacks, but while lowering the original accuracy due to the change in classification paradigm~\cite{tsipras2018there,MadryMSTV17}. Some extensions to adversarial training were implemented to improve the classification accuracy, including the virtual adversarial training using smoothing and LDS gradient, by Miyato~\etal~\cite{miyato2015distributional}, and ensemble adversarial training, which improves the quality of adversary training sample, by Tramer~\etal~\cite{TramerKPBM17}. Transfer learning techniques, such as distillation, were used by Papernot~\etal~\cite{PapernotM0JS16}, to improve the robustness of the aforementioned approach.

\begin{figure*}[t]
\centering
\includegraphics[width=0.95\textwidth]{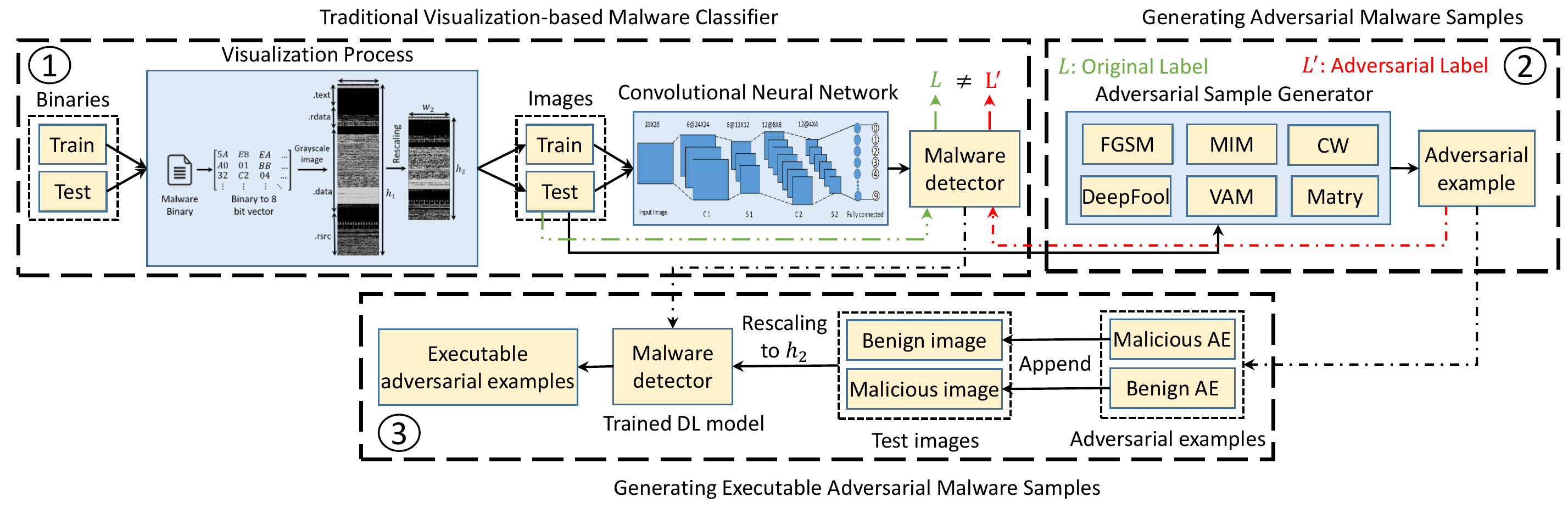}
\caption{General structure of the visualization-based generic adversarial attacks. As shown, it consists of three modules. Module 1 is for the visualization-based malware detector, module 2 is for generating adversarial samples based on generic adversarial learning algorithms. Note that the AEs generated at module 2 are not necessarily executable. Thus, to maintain the executability of the crafted AE, the generated AEs from 2 are being appended to the original test images in the third module.}
\label{fig:Malware_Practical}\vs{3}
\end{figure*}

\section{Methods and Techniques}
\label{sec:methedology}
There are several studies that incorporate deep learning-based classifiers for detecting malicious binaries from benign ones. In addition, it has been shown that deep learning models are not immune to adversarial attacks, and can be easily fooled. Although the concept of adversarial machine learning is an active research area, there are very few research works that have investigated the impact of adversarial attacks on the deep learning-based malware detection systems~\cite{liu2019atmpa}. Therefore, the key goals of our study is to investigate the impact of adversarial attacks on the performance of deep learning-based malware detector, as well as generating adversarial malware samples that are executable. In particular, we aim to generate adversarial samples that can fool the deep learning model such that malware binaries are classified as benign, and vice versa.

To do so, we design three different configurations: an adversarial attack configuration, a defense configuration, and an executability configuration. The first configuration utilizes several existing attack methods---previously unexplored for this problem domain---to generate adversarial binary samples. The second configuration explores potential defense strategies to increase the robustness of deep learning-based malware detectors against adversarial examples. In the executability configuration, we focus on generating adversarial samples that are not only successful in fooling the classifier, but also are executable. In the following, we will describe step-by-step aspect of these configurations.

\subsection{Generic Adversarial Attacks} \label{sec:method_attacks}

To investigate off-the-shelf adversarial attacks on malware detection, we utilized five generic algorithms capable of generating AEs. The primary goal of this configuration is to generate AEs that successfully fool the traditional visualization-based malware detection designed using deep learning networks. This attack configuration consists of two modules, the traditional visualization-based malware detection system (\ding{172} in~\autoref{fig:Malware_Practical}) and adversarial example generator module (\ding{173} in~\autoref{fig:Malware_Practical}), reviewed in the following.

\subsubsection{Visualization-based Malware Detection System} This module consists of a visualization and deep learning sub-modules. The visualization sub-module transforms the binary code of each (potential) malware sample into a grayscale image, while the deep learning sub-module, which is based on a convolutional neural network with multiple layers, is designed and trained on a training set of grayscale images and evaluated using the test images for tuning. Note that this model builds the baseline model for malware detection, which will be mainly used for adversarial binary generation in the subsequent modules.

\subsubsection{Adversarial Example Generator} 
The main responsibility of this module is to create AEs using off-the-shelf adversarial machine learning algorithms. The adversarial sample generator module, shown in Figure~\ref{fig:Malware_Practical} takes the trained model and test set of grayscale images as inputs and generates adversarial binaries using one of five different algorithms: the Fast Gradient Sign Method (FGSM), due to Goodfellow \etal~\cite{Goodfellow2015Explaining}, Carlini and Wagner method (C\&W), due to Carlini and Wagner~\cite{Carlini017}, DeepFool, due to Moosavi-Dezfooli \etal~\cite{Moosavi-Dezfooli16}, Momentum Iterative Method (MIM), due to Dong \etal~\cite{dong2018boosting}, and Projected Gradient Descent (PGD), due to Madry \etal~\cite{MadryMSTV17}. We briefly describe each attack method in~\autoref{app1}.

\subsubsection{Shortcomings} This attack configuration assumes the availability of an intermediary state, rather than the malware binary, where binaries are already transformed into grayscale images (visualized) accessible by the adversary, and the adversary tries to force the trained DL model into misclassification. While this constitutes a major shortcoming of this configuration, it is not the only one: another shortcoming is that the generated AEs in this configuration are not guaranteed to be executable since the applied perturbation into the grayscale image may result in corrupted binary code. 

In conclusion, this configuration is well-suited to cases where the detection design is based on the visualization approach alone, where malware binary code are readily  transformed into grayscale images. This configuration, however, will not be effective with end-to-end detection (i.e., where the detector starts with the binary, does the transformation via visualization, then learning and detection).

\subsection{Defense Configuration} \label{sec:method_defense}
The goal of the defense is to improve the robustness of the malware detection system against AEs. 
Although there are multiple defensive algorithms against AEs, such as defensive distillation~\cite{PapernotM0JS16} and adversarial training~\cite{Goodfellow2015Explaining,szegedy2014intriguing,papernot2017practical}, none of them is completely effective. Nonetheless, in this study we utilized adversarial training to improve the robustness of our malware detection system, as it is the most successful adversarially robust model so far~\cite{tsipras2018there}. The basic idea of adversarial training is to inject AEs into the training procedure to increase the robustness of the model~\cite{tsipras2018there}, where the goal is to solve the following adversarial empirical risk minimization:
\begin{equation*}
    \min_{\theta }\ \textup{E}_{(x,y)\sim \widetilde{D}} \left [max_{\delta \epsilon S } \ L(x+\delta, y; \theta) \right ],
\end{equation*}
where $x$ is the input image, $y$ is the model output, $\theta$ is the model parameters, and $\delta$ is a small perturbation. 

The steps taken to defend against the adversarial samples are shown in~\autoref{al:TOC_methedology}. The algorithm takes $Dataset_{org}$, $Labels$, and $CNN_{tr}$ as inputs, generates adversarial samples against $CNN_{tr}$ using the previous five attacking methods and train a new method over the combination of the original dataset and generated adversary examples. Each sample in the original dataset will be represented with additional five adversary samples each from different attack. This will increase the robustness of the model as the decision will be related to the core features that preserve the information, rather than being affected by the change of features that is not fundamentally meaningful~\cite{tsipras2018there}, which may lead to overfitting.

\subsection{COPYCAT} \label{sec:method_executable}

In a nutshell, generating adversarial samples while affecting the executability of the application is counter-intuitive, and is an easy way to infer whether a piece of software is malicious or benign (by attempting to execute it). Moreover, altering a malware sample while not paying attention to modifications in the malicious piece of code (by, for example, overriding it), while may fool the classifier, goes against the whole point of adversarial example generation from a practical standpoint. Therefore, we propose two approaches to generate samples that 1) fool the classifier, and 2) maintain the executability of the original binary (malicious or benign). 

The key idea in this work is inspired by the work of Cha \etal's~\cite{ChaPBL10}, where binaries of an application specified by a certain architecture are padded into an application from another architecture to change the behavior and application logic depending on the system used for running it. As such, we consider the machine learning model as an additional system on which a sample needs to operate, along with the actual system on which the binary will run, and customize our sample to operate successfully on both as intended. Note that, and unlike the work of Cha \etal's~\cite{ChaPBL10}, our program sizes do not need to be identical  in size: we have the advantage of constructing a sample of a minimal size only to fool the classifier, while not altering the actual sample for a guaranteed executability. 

\subsubsection*{Executable Adversarial Examples} \ours contributes two methods for executable adversarial samples generation:  

\begin{itemize}[leftmargin=*]

\item {\bf AE Padding.} Our first method is called ``AE Padding'', in which \ours generates an AE $x'$ for an original sample $x$ using one of the five previous attack methods. Then, we convert $x'$ to an adversarial image with the same dimensions of the original sample image, and append the adversarial image to the end of the original sample image. Appending pixels at the end is equivalent to appending the code into a non-reachable code area, which will not affect the executability of the program but will change the output pattern of the visualization, and likely leading to a misclassification (which needs to be confirmed). The general architecture of this configuration of \ours is shown in~\autoref{fig:Malware_Practical}, where the third module in \ding{174} is responsible for appending AEs, generated in \ding{173}, to the end of the original sample image. Note that the appended images were not necessarily executable, whereas the output of the third module is executable. 

\item {\bf Sample Injection.} The second method is called ``Sample Injection'', in which we inject the binaries of a sample from the targeted class into an unreachable area of the binaries (code) of the attacked sample, usually at the end of the code. This process will ensure the executability of the original binary while changing the pattern of the visualization, leading to a model misclassification. 
\end{itemize}

\BfPara{Concurent Work} Two other concurrent studies~\cite{KolosnjajiDBMGE18, kreuk2018deceiving} proposed similar approaches to ours. However, unlike those concurrent studies which produced untargeted misclassification attacks, \ours produces both targeted and untargetted misclassification. In other word, \ours can force the model to label the original input as any other desired class. This is a particularly important objective in malware family classification applications, since classification, for example, is used for attribution, designing the correct remedies, and risk management procedures, etc. Moreover, although a minor contribution, we explored and confirmed that our approach generalized equally well to IoT malware. Moreover, we show that \ours is able to achieve an untargeted misclassification rate of 98.90\%, compared to the untargetted classification results of 60\% in the concurrent published work~\cite{KolosnjajiDBMGE18}. Finally, \ours was able to achieve a targeted misclassification rate of 73.50\% and 96.5\% for Windows and IoT samples, respectively, which are promising. We use a CNN model with a general architecture that can accommodate multiple types of malware, unlike MalConv~\cite{KolosnjajiDBMGE18}.

\begin{algorithm}[t]
    \SetKwInOut{Input}{Input}
    \SetKwInOut{Output}{Output}
    Function \textbf{AdvTraining} $(\texttt{$D_{o}$, $L_o$,  $M_{o}$})$\;
    \Input{$D_{o}$, $L_o$,  $M_{o}$}
    \Output{$M_{A}$}
    
    ${D_{t}}\gets$ $D_{o}$
    
    ${L_{t}}\gets$ $L$
    
    \For{attack $IN$ adversary_attacks}{
      $AEs \gets attack.generate_{}(M_{o})$
      
      $D_{t} \gets D_{t} \cup AEs$
      
      $L_{t} \gets L_{t} \cup L_o$
      }
    $M_{A} \gets$ New Model
    
    $M_{A}$.train($D_{t}$ , $L_{t}$)
    
    \Return {$M_{A}$}
    
    \caption{Sample algorithm for adversarial training as a defence against adversarial attacks. Here, the model ($M_o$) trained over the original dataset ($D_o$) and the corresponding labels ($L_o$) are inputs of the function ($AdvTraining$) and adversarially robust model ($M_a)$ is its output.  }
    \label{al:TOC_methedology}
    \end{algorithm}

\section{Evaluation and Discussion}\label{sec:Results}
We evaluate the performance of \ours for generating executable AEs through comprehensive experiments. To do so, we assume that the adversary aims to conduct both targeted and untargeted attacks, with  white-box access to the model. We evaluate the performance of \ours against two set of datasets, namely Windows and IoT binaries. In the following, we review the datasets, the experimental setup, the results and finding, and a discussion.

\subsection{Dataset}
We used two datasets: Microsoft Windows malware binaries and Internet of Things malware binaries. We used those two types of malware for the following reason. First, while Windows malware is popular, IoT malware is on the rise, and deserves consideration in evaluations. Second, because machine learning algorithms outcomes depend on the used dataset, having multiple datasets would be better than one. Finally, we note that no prior work used such emerging IoT malware dataset, although our work can be easily applied to other malicious binaries, such as Android applications as well. To this end, in the following we review the used datasets. 
\begin{itemize}[leftmargin=*]
\item {\bf Windows Malware Binaries. }
To carry out our experiment on representative Windows malware samples and facilitate reproducibility of results, we used open source malware dataset, called BIG 2015. BIG 2015 is introduced in the Kaggle Microsoft Malware Classification Challenge~\cite{ronen2018microsoft}. The dataset consists of 10,678 labeled malware samples classified into nine different malware families, and 1,000 benign samples randomly selected from valid executables of Windows 10 binaries. \autoref{WindowsDist} shows the distribution of the samples across the nine malware families.

\item {\bf IoT Malware Binaries.} We select IoT malware for their prevalence, as they are very popular for the increasing use of IoT devices, their flexibility (Linux applications) and portability (developed for multiple platforms). For our IoT binaries, we obtained a set of 2,899 randomly selected IoT malware samples in binary format from the IoTPOT project~\cite{PaSYMKR16}. In addition, we collected a set of 207 benign IoT binaries, for a total of  3,106 IoT binaries. 
\end{itemize} 

\begin{table}[t]
\centering
\caption{Family-level distribution of Windows malware.}
\label{WindowsDist}
\begin{tabular}{l|c|c}
\toprule
\multirow{2}{*}{Malware Family} & \multicolumn{2}{c}{Number of Samples}\\
                      &  \# & $\%^{1}$\\
\midrule
Ramnit &  1,534  &   13.14\%  \\ 
Lollipop  & 2,470  &   21.15\%   \\ 
Kelihos ver3 & 2,942  &   25.19\% \\ 
Vundo  & 451  &   3.86\%   \\ 
Simda  & 41  &   0.35\%   \\ 
Tracur  & 685  &   5.87\%   \\ 
Kelihos ver1  & 386  &   3.30\%  \\ 
Obfuscator.ACY  & 1,158  &   9.91\%  \\ 
Gatak  & 1,011  &   8.65\%  \\ 
Benign Sample  & 1,000  &   8.56\%  \\ 
 \midrule
 Total &  11,678 & 100\% \\
\bottomrule
\end{tabular}
\end{table}

\begin{table}[t]
\centering
\caption{Distribution of IoT malware samples across families (labels assigned by antivirus scanners). $\%^{1}$ refers to the representation of families among malware samples.}
\label{ArchDist}
\begin{tabular}{l|c|c}
\toprule
\multirow{2}{*}{Arch} & \multicolumn{2}{c}{Number of Malware}\\
                      &  \# & $\%^{1}$\\
\midrule
Gafgyt &  2,609  &   89.99\%  \\ 
Mirai  & 185  &   6.38\%   \\
Tsunami & 67  &   2.32\% \\
Singleton  & 32  &   1.10\%   \\
Hajime  & 6  &   0.21\%   \\ 
 \midrule
 Total &  2,899 & 100\% \\
\bottomrule
\end{tabular}
\end{table}

\autoref{ArchDist} shows the distribution of the malware samples across families, the variety in malware IoT samples helps with creating more robust models exposed to more patterns and behaviors which increases the decision confidence of the classifier. \autoref{IoT_Training} shows the labels used in the training process of the DL model, summing up the number of benign and malicious samples outlined above.

\begin{table}[t]
\centering
\caption{IoT dataset: malware and benign binaries.}
\label{IoT_Training}
\begin{tabular}{l|c|c}
\toprule
\multirow{2}{*}{Label} & \multicolumn{2}{c}{Number of Samples}\\
                      &  \# & $\%^{1}$\\
\midrule
Benign &  207  &   6.66\%  \\ 
Malware  & 2899  &   93.34\%   \\
 \midrule
 Total &  3,106 & 100\% \\
\bottomrule
\end{tabular}
\end{table}

\subsection{Experimental Setup}

For a transparent evaluation, in the following we describe the experimental setup we use for generating the adversarial examples.

\begin{itemize}[leftmargin=*] 
\item {\bf Algorithms implementation.} All attack methods are implemented using Cleverhans~\cite{papernotCGFFMHJKS2016}, a python library with state-of-the-art implementations for various adversarial methods. 

\item {\bf Parameters.} Each attack method has different parameters. We conducted several experiments with different values for such parameters to achieve a high misclassification rate. For example, DeepFool has two parameters to be tuned, the overshooting value and the number of iterations, which were tuned to $100$ and $0.05$, respectively, to achieve  96.43\% accuracy with the Windows malware model. The parameters of each attack method are reported in~\autoref{tab:Attack_Parameters} (for both datasets), while the obtained results for other attack methods are reported in~\autoref{tab:windows_methods} and~\autoref{tab:IoT_methods} for the Windows and IoT datasets, respectively. 
\item {\bf Evaluation  system.} All experiments are conducted using Python 3.6 running over Ubuntu 16.04 and using a system of i5-8500 CPU operating at 3.00GHz, with 32GB DDR4 RAM, 512GB SSD, and NVIDIA GTX980 Ti graphical processor unit (GPU).
\end{itemize}

\begin{table}[t]
\centering
\caption{Parameters used for each adversarial attack methods. Here $\epsilon$ refers to the distortion parameter, L.R. refers to learning rate, \# Iter. refers to the number of iterations, and O. refers to overshooting parameter in DeepFool method. The sign (--) means the parameter is not applicable to the method.}
\begin{tabular}{|c|c|c|c|c|}
\hline
Method & $\epsilon$ & \# Iter. & L.R. & O. \\
\hline
FGSM & 0.3 & - & - &  - \\ \hline
C\&W & - & 100 & 0.1 &  - \\ \hline
DeepFool & - & 100 & - &  0.05 \\ \hline
PGD & 0.3 & 250 & - &  - \\ \hline
MIM & 0.3 & 250 & - &  - \\ \hline

\end{tabular}\label{tab:Attack_Parameters}
\end{table}

\subsection{Evaluation Metrics}
We use three evaluation metrics: the mislcassification rate (MR), perturbation magnitude (PM), and running time (RT). MR is used as an indication of the success rate for fooling the detection system. When the predicted label $\argmax_k P(y_k|\bar{x}_i)$ for the adversarial sample 
$\bar{x}_i$ is not same as the correct class label $y_i$ of the original sample $x_i$, this results in misclassification, and the rate is calculated as:
$$
\frac{1}{n}\sum_i I(\argmax_k P(y_k|\bar{x}_i) \neq y_i). 
$$

We use $L_p$, which  measures the PM by $p\mbox{--}norm$
distance as:
\[ \left \| \delta \right \|_p = \left ( \sum_{i=1}^{n} \left \| \bar{x}_i - x_i \right \|^p \right )^{\frac{1}{p}} \]
For $p\mbox{--}norm$, studying the $L_0$, $L_2$ and $L_\infty$ is very common~\cite{Carlini017}. In this study, we measure the count of changes in the adversarial example compared to the original sample and $L_2$ distance of the adversarial example and the original one. Note that these two metrics indicate magnitude of the perturbations. Finally, we measure the running time (RT) required to generate adversarial examples (in seconds) as a metric to compare the complexity of various attacks. 

\subsection{Results and Discussion}
 In order to provide better perspective of our findings, by following the main key contributions highlighted in our methodology, this section is broken down into attack configuration (\autoref{sec:Attack_Configuration}), defense configuration (\autoref{sec:Defense_Configuration}), and executable configuration (\autoref{sec:Executable_Configuration}). For each of these configurations, we discuss the results for both of the Microsoft Windows and IoT malware datasets.

\subsubsection{Generic Attacks Configuration} \label{sec:Attack_Configuration}

Our main goal is to generate adversarial samples that are able to fool the DL-based malware detector. Note that this configuration is based on applying the perturbations on the images generated from malware binaries, thus there is no guarantee regarding the executability of the crafted AEs. 

\BfPara{Windows} \label{attacks_metrics}
We generate adversarial examples using five generic adversarial attack methods mentioned in~\autoref{sec:method_attacks}. ~\autoref{tab:windows_methods} shows the results for the different attack methods in terms of misclassification rate, perturbation magnitude, and running time. 
As shown in~\autoref{tab:windows_methods}, PGD achieved the highest misclassification rate of 99.63\%, while perturbing 89.14\% of the pixels. However, PGD is expensive in term of its running time. On the other hand, FGSM and C\&W are the top two methods with the lowest running times, while reaching reasonable misclassification rate of 99.08\% and 99.45\%, respectively. 

One interesting observation from~\autoref{tab:windows_methods} is that all of these generic adversarial attack algorithms change a huge portion of the original image's pixels (e.g., more than 71\%), which in turn correspond to parts of the original binaries. Thus, it can be easily expected that the applied perturbation on the grayscale images may result in a corrupted binary that will not be executable. 

\begin{table}[t]
\centering
\caption{Generating adversarial examples for Windows binaries using generic adversarial attack methods. }
\scalebox{0.9}{
\begin{tabular}{|c|c|c|c|c|c|}
\hline

\multirow{2}{*}{Method} &  \multirow{2}{*}{MR (\%)} & \multicolumn{2}{c|}{Pixels} & \multirow{2}{*}{$L_2$ Dist.} & \multirow{2}{*}{RT (s)}\\
 & &  \# & \% & & \\ 
\hline
FGSM & 99.08         &  8,594 & 83.92   & 25.86   & \textbf{112.10}\\ \hline
PGD & \textbf{99.63}        &  9,128 & 89.14  & 23.11  & 1,339.60  \\ \hline
C\&W  & 99.45         & \textbf{7,316}  & \textbf{71.44}  & 4.09   & 2,071 \\ \hline
MIM  & 99.54          &   8,732 & 85.27  & 25.08  & 6,875 \\ \hline
DeepFool & 99.45         & 7,750  & 75.68   & \textbf{3.23}    & 15,545.70 \\ \hline
\end{tabular}\label{tab:windows_methods}}
\end{table}

\BfPara{IoT} 
The same set of attacks and evaluation metrics as in~\autoref{attacks_metrics} are applied on the IoTPOT dataset.  As such, the results with the IoTPOT dataset are shown in~\autoref{tab:IoT_methods}. As shown, we can observe that all adversarial attack methods achieved a misclassification rate of 100\%, except for the FGSM method. Here the C\&W method is shown to outperform other methods considering the $L_2$ distance between the generated AEs and original inputs, although with a larger number of changed pixels. In particular, we notice that the number of changed pixels varies from 87.22\% in C\&W to 93.33\% in the FGSM approach. The huge number of changed pixels in the generic adversarial  approaches may make the crafted adversarial binaries unexecutable, which is not desirable in effective evasion techniques for software (that is supposed to be executed).

\BfPara{\textcolor{blue!70!black}{Executability}} We note {\em all} adversarial examples resulting from both malware datasets, using all methods above were unexecutable.

\begin{table}[t]
\centering
\caption{Generating adversarial examples for IoT binaries using different adversarial attack methods}
\scalebox{0.9}{
\begin{tabular}{|c|c|c|c|c|c|}
\hline
\multirow{2}{*}{Method} &  \multirow{2}{*}{MR (\%)} & \multicolumn{2}{c|}{Pixels} & \multirow{2}{*}{$L_2$ Dist.} & \multirow{2}{*}{RT (s)}\\
 & &  \# & \% & & \\ 
\hline
PGD & \textbf{100}        &  9,452 & 92.30  & 23.36  & 3,979.80  \\ \hline
FGSM & 93.19         &  9,558 & 93.33   & 26.98   & \textbf{51}\\ \hline
C\&W  & \textbf{100}         & \textbf{8,932}  & \textbf{87.22}  & \textbf{2.44}   & 975.30 \\ \hline
MIM  & \textbf{100}          &  9,296 & 90.78  & 24.77  & 3,926.50 \\ \hline
DeepFool & \textbf{100}         & 9,258  & 90.41   & 3.68    & 1,798.10 \\ \hline
\end{tabular}\label{tab:IoT_methods}}
\end{table}

\subsubsection{Defense Configuration} \label{sec:Defense_Configuration}
Defense configuration aims to improve the robustness of the DL-based malware detection systems against various generic adversarial attack methods.  

\BfPara{Windows}
We implement the adversarial training method to increase the robustness of the DL-based malware detector against generic adversarial attack methods. To do so, we trained a model using the original inputs and the corresponding adversarial examples, which were generated using the aforementioned generic algorithms. The misclassification rates before and after the adversarial training for all adversarial attack methods are shown in~\autoref{tab:defence_AT}. As it can be seen from~\autoref{tab:defence_AT}, training over adversarial examples improves the robustness of the neural network against adversarial learning attacks. For instance, the evasion rate has dropped significantly from 99.08\% into 8.68\% (which is more than 90\% improvement on the robustness of the model) for FGSM in Windows binaries after adversarial training. We observe similar pattern for other adversarial attack approaches. 

\begin{figure}[t]
\centering
\includegraphics[width=0.4\textwidth]{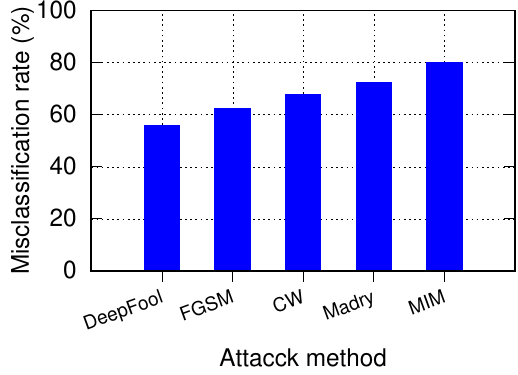}
\caption{Misclassification rate of benign executable files due to appended malicious adversarial samples. We found that the misclassification rate is highly correlated with the \# of distorted pixels and the $L_2$ distance.}\vs{5}
\label{fig:Win_AE_Padding}
\end{figure}

\BfPara{IoT}
We implement the same defense for the IoT malware detector by training a model on the IoT dataset and generating the corresponding adversarial examples. The misclassification rate before and after adversarial training for all adversarial attack methods are shown in~\autoref{tab:defence_AT}.  As it can be seen from~\autoref{tab:defence_AT}, training over adversarial examples improves the robustness of the neural network against the adversarial learning attacks. For instance, the robustness of the DL-based IoT malware detection system has improved more than 99\% for the FGSM approach in IoT binaries after adversarial training. The same pattern is observed for other methods.

\begin{table}[t]
\centering
\caption{Misclassification rate on Windows and IoT before and after applying adversarial training. * indicates to the misclassification rate after applying adversarial training.}
\begin{tabular}{|c|c|c|c|c|}
\hline
\multirow{2}{*}{Method} & \multicolumn{2}{c|}{Misclassification (\%)} &\multicolumn{2}{c|}{Misclassification* (\%)} \\
& Windows & IoT & Windows & IoT \\
\hline
FGSM & 99.08 & 93.19 & 8.68 &  0.80 \\ \hline
C\&W & 99.45 & 100 & 26.12 &  15.76 \\ \hline
DeepFool & 99.45 & 100 & 83.24 &  97.43 \\ \hline
PGD & 99.63 & 100 & 27.80 &  5.47 \\ \hline
MIM & 99.54 & 100 & 77.76 &  4.98 \\ \hline
\end{tabular}\label{tab:defence_AT}
\end{table}

\subsubsection{COPYCAT} \label{sec:Executable_Configuration}
The results of our proposed approach, \ours, for generating executable adversarial binaries are reported as followed.  
\BfPara{Windows}
 In this paper, we introduced AE padding and sample injection. ~\autoref{fig:Win_AE_Padding} shows the results of the AE padding method, by appending the adversary example, generated using generic adversarial atatck methods, to the binaries of the original executable sample. As it can be seen, we achieved the highest misclassification of 79.9\% using MIM as our adversarial example generator. \autoref{tab:windows_sampleMalwareInjection} and ~\autoref{tab:windows_sampleBenignInjection} show the results of the sample injection method, by injecting  benign sample binaries at the end of malware sample binaries (\underline{M2B}), and vice versa (\underline{B2M})---when we inject a benign sample into a malware sample we are in fact converting a malware sample into a benign sample (M2B). In the first case, we achieved an overall misclassification rate of 98.9\%, while the misclassification rate of M2B was 73.5\% by injecting a benign sample with 4.4MB size. In the second case, we achieve a misclassification rate of B2M (regardless of the class) of 98.6\% with a malware sample of size 11.5MB.

\begin{table*}[t]
\parbox{.19\linewidth}{
\centering
\caption{\normalfont Size vs. MR of B2M (Windows)}

\begin{tabular}{|c|c|}
\hline
Malware size & MR(\%) \\
\hline
0.82 MB  & 82.1   \\ \hline
1.10 MB &  88.5 \\ \hline
2.20 MB & 96.4 \\ \hline
11.50 MB &  \textbf{98.6} \\ \hline

\end{tabular}\label{tab:windows_sampleMalwareInjection}
}
\hfill
\parbox{.19\linewidth}{
\centering
\caption{\normalfont Size vs. MR of M2B (IoT). }
\begin{tabular}{|c|c|}
\hline
Benign size & MR (\%) \\
\hline
1.0 MB  & 67.5   \\ \hline
1.9 MB &  76.7 \\ \hline
3.6 MB & 95.5 \\ \hline
11.5 MB &  \textbf{96.5} \\ \hline

\end{tabular}\label{tab:IoT_sampleBenignInjection}
}
\hfill
\parbox{.19\linewidth}{
\centering
\caption{\normalfont Size vs. MR of B2M (IoT).}
\begin{tabular}{|c|c|}
\hline
Malware size & MR (\%) \\
\hline
256 KB  & 56.50   \\ \hline
553 KB  & 92.27   \\ \hline
1.1 MB &  95.17 \\ \hline
2.4 MB  & \textbf{95.65} \\ \hline
\end{tabular}\label{tab:IoT_sampleMalwareInjection}
}
\hfill 
\parbox{.19\linewidth}{
\centering
\caption{\normalfont Transferability: Size vs. MR of B2M (Win).}

\begin{tabular}{|c|c|}
\hline
Malware size & MR(\%) \\
\hline
0.82 MB  & 90.09   \\ \hline
1.10 MB &  94.06 \\ \hline
2.20 MB & 78.21 \\ \hline
11.50 MB &  \textbf{100} \\ \hline

\end{tabular}\label{tab:GeneralizationWindows_sampleMalwareInjection}
}
\hfill
\parbox{.19\linewidth}{
\centering
\caption{\normalfont Transferability of M2B (IoT). }
\begin{tabular}{|c|c|}
\hline
Benign size & MR (\%) \\
\hline
1.0 MB  & 56.00   \\ \hline
1.9 MB &  70.50 \\ \hline
3.6 MB & 36.75 \\ \hline
11.5 MB &  \textbf{80.00} \\ \hline

\end{tabular}\label{tab:GeneralizationIoT_sampleBenignInjection}
}
\end{table*}

\begin{table}[t]
\centering
\caption{\normalfont Size vs. MR of M2B and overall (Windows).}
\begin{tabular}{|c|c|c|}
\hline
\multirow{2}{*}{Benign Size} & \multicolumn{2}{c|}{Misclassification (\%)} \\ 
& Overall & Benign \\
\hline
1.2 MB &  76.2 &  56.1 \\ \hline
1.4 MB & 90.2   &  37.8  \\ \hline
3.5 MB &  89.3 & 63.0 \\ \hline
4.4 MB & \textbf{98.9}   & \textbf{73.5}\\ \hline
\end{tabular}\label{tab:windows_sampleBenignInjection}
\end{table}

\BfPara{IoT} We show results for the sample injection method, which consists of injecting benign binaries into malware and vice versa.~\autoref{tab:IoT_sampleBenignInjection} shows the results of injecting benign binaries into malware samples, by using a benign sample of size 11.5MB, we achieve a misclassification of 96.5\%. Whereas~\autoref{tab:IoT_sampleMalwareInjection}  shows the results of injecting malware binaries into benign samples, achieving a misclassification accuracy of 95.65\% using a malware of size 2.4MB.

\BfPara{\textcolor{blue!70!black}{Executability}} {\em All} adversarial examples that resulted from both malware datasets, using the two methods above, were executable.

\subsection{Generalization with Transferability}
To investigate the generalization of the generated adversarial binaries using \ours, several experiments are conducted. 
To do so, we train another DL model with the following network structure: we use three consecutive fully connected layers of size 64 connected to the input vector, followed by a dropout with a probability of 0.5. Similarly, the  output  of  the  dropout function  is  fully  connected  with  another  softmax dense layer. Throughout our experiments we found that the injection of malware samples can force this Deep Neural Network (DNN)-based model to misclassification, with a similar performance to that of the CNN-based malware detection system. The results of the benign-to-malware misclassification attack are shown in~\autoref{tab:GeneralizationWindows_sampleMalwareInjection}, where it is concluded that the generated adversarial binaries are general, and can be used to possibly fool other models with different structures. For instance, the sample injection configuration of \ours is able to force the DNN-based malware detection system to misclassify benign samples to malicious  with a varying misclassification rate from 78.21\% to 100\% for samples injected with size of 2.2MB and 11.5MB, respectively. In addition, we found that the injection of a benign sample into a malicious sample can result in an evasion rate of up to 89.28\% in a DNN-based malware detection system, as the detailed results are shown in~\autoref{tab:Transferability_BenignInjection}.

We also observed a similar set of results for the IoT dataset, as shown in ~\autoref{tab:GeneralizationIoT_sampleBenignInjection} and ~\autoref{tab:GeneralizationIoT_sampleMalwareInjection} for benign and malicious sample injection, respectively. For example, we were able to achieve a misclassification rate of 80\% and 97.58\% by injecting a benign and malicious file, respectively, in DNN-based IoT malware detection. 

\begin{table}
\parbox{.39\linewidth}{
\centering
\caption{\normalfont Transferability: Size vs. MR of B2M (IoT).}
\begin{tabular}{|c|c|}
\hline
Malware size & MR (\%) \\
\hline
256 KB  & 85.02   \\ \hline
553 KB  & 82.60   \\ \hline
1.1 MB &  95.17 \\ \hline
2.4 MB  & \textbf{97.58} \\ \hline
\end{tabular}\label{tab:GeneralizationIoT_sampleMalwareInjection}
}
\hfill
\parbox{.59\linewidth}{
\centering
\caption{\normalfont Transferability: Size vs. MR of M2B and overall (Windows).}
\begin{tabular}{|c|c|c|}
\hline
\multirow{2}{*}{Benign Size} & \multicolumn{2}{c|}{Transferability (\%)} \\ 
& Overall & Benign \\
\hline
1.2 MB &  37.5 &  23.7 \\ \hline
1.4 MB & 72.4   &  18.1  \\ \hline
3.5 MB &  66.0 & 42.3 \\ \hline
4.4 MB & \textbf{89.3}   & \textbf{34.4}\\ \hline
\end{tabular}\label{tab:Transferability_BenignInjection}
}
\end{table}

\subsection{Size Trade-off}
Injection method's outcomes (misclassification) depend on the size of the injected sample, due to the behavior of NN in general. Neural networks are well known for recognizing patterns, and increasing the size of the injected sample will increase the portion of sample's pattern against the overall layout. In general, a larger  injected sample results in a higher effect on misclassification.

The attacks discussed in~\autoref{attacks_metrics} do not guarantee the executablity of the generated sample, as each pixel in the visualization corresponds to some of the data in the original sample. Changing one bit (or pixel) may corrupt the sample, leading to unexecutable adversarial example. Our proposed methods preserve the functionality and executability by padding data into unreachable point of the application, which is usually after the exit code. These methods will not affect the application in any way except by increasing its size, which we consider as a trade-off cost of executability.

\section{Related Work}\label{sec:related}
Machine and deep learning algorithms for malware detection have been actively pursued. For example, 
malware visualization is a technique that transforms binaries into images, which are then used to build machine/deep learning-based malware detection systems~\cite{lee2011study, ZhangQYOH16}. Cui \etal~\cite{CuiXCCWC18} introduced a malware detection method using deep learning, and by transferring the malicious code into grayscale images. They achieved an accuracy rate of 94.5\% on the Vision Research Lab dataset~\cite{nataraj2011malware}, claiming better accuracy rate and execution time compared to static~\cite{kang2015detecting} and dynamic feature analysis~\cite{Al-DujailiHHO18}.  Similarly, Ni \etal~\cite{NiQZ18} proposed a malware detection system, called MCSC, that is build over 10805 grayscale images consisting of nine different malware families. Su \etal~\cite{SuVPSFS18} presented a lightweight IoT malware classifier that works based on visualization and CNN model, and achieves a detection rate of 94\% over goodware and DDoS malware classification tasks~\cite{pa2015iotpot}. Fu \etal~\cite{FuXWLS18} presented a fine-grained malware detector, through RF, KNN, and SVM models, that was trained over colored images generated from malware binaries. Their method was able to achieve a malware detection accuracy rate of 97.47\%, and family classification accuracy rate of 96.85\% using random forest model. Moreover, Fan \etal~\cite{FanHZYA18} presented a Metagraph2vec based malware detection system. Hou \etal~\cite{HouYSA17} investigated android malware detection system based on structured heterogeneous information network. 

Arp \etal~\cite{ArpSHGR14} presented a light-weight method, DREBIN, for detecting malicious android applications based on machine learning models trained over features extracted from static analysis. McDermott \etal~\cite{mcdermott2018botnet} introduced  a Mirai detector based on deep learning, which could distinguish Mirai botnet traffic from normal IoT IP camera traffic with 99\% accuracy. Pajouh \etal\cite{haddadpajouh2018deep} explored the potential of using recurrent neural networks to detect ARM-based IoT malware, achieving detection accuracy  of 98.18\% using LSTM structures to analyze OpCode sequences. Furthermore, Wang \etal~\cite{WangGZOXLG17} proposed an adversary resistant neural network with an application to malware detection based on feature nullification. 

Although addressing a different problem, and using a different approach, \ours is inspired in part by Cha \etal's~\cite{ChaPBL10}, where they propose the idea of platform-independent programs. Moreover, the approach in this work is in line with although concurrent to the work of Kolosnjaji \etal ~\cite{KolosnjajiDBMGE18} and kreuk \etal~\cite{kreuk2018deceiving}. However, \ours is different in multiple aspects, as highlighted earlier; \eg performing both untargeted and targeted misclassification, tackling the visualization-based malware detection approach, providing higher evasion rate, and targeting both windows and IoT malware. We claim that \ours is able to acheive high attack success rate while generating executable adversarial binaries, unlike the work of Liu \etal ~\cite{liu2019atmpa}, where the main focus is maximizing the transferability regardless of the executability of the generated binaries.

\section{Limitation and Future Work}\label{sec:Limitation}

We have explored two methods to generate adversarial examples that fool the visualization-based malware detection system, while preserving executability and functionality of the binaries. However, both of the proposed methods require appending data into the original sample. Generalizing this over multi-class classifiers, requires padding of at least one AE of each target class to ensure targeted misclassification. For example, to  support a 10-class classifier would require multiplying the size of the original sample by approximately 10 times. This, in turn, constitutes the main shortcoming of the AE padding method. In addition, our results indicate that the performance of the sample injection method positively correlates with the size of the injected sample. Furthermore, the adversarial training method depends on the injected AEs similar to the ones used by the adversary, which may not be the case in practice. It is computationally expensive and almost impossible to find adversarial inputs by all attack models, which results in that the adversarial training is not-adaptive to unknown attacks.

\section{Conclusion}\label{sec:Conclusion}

Recent research shows that deep learning is vulnerable to adversarial examples, whereby an adversary can manipulate the input of a deep learning model in such an unnoticeable way by adding minimal perturbation to the input to generate a desired output, \ie misclassification. In this work, we have studied several generic adversarial attacking methods and incorporated them into deep learning malware detection, showing a great success in fooling classifiers. To allow misclassification while maintaining executability, a shortcoming of state-of-the-art, we proposed AE padding and sample injection, two new adversarial methods to manipulate the output of the deep model while guaranteeing the functionality and executability of the generated sample. By designing three configurations; an adversarial attack configuration, a defense configuration and an executability configuration, we have evaluated the performance of the proposed attacks through extensive experiments. The executable configuration is capable of generating adversarial examples that successfully force the deep learning-based model to a misclassification rate of 98.9\% and 96.5\% on Windows and IoT malware datasets, respectively, while preserving executability and maintaining the same functionality of the original malware.

 \BfPara{Acknowledgements} This work is supported by NSF grant CNS-1809000, NRF grant 2016K1A1A2912757, NVIDIA GPU Grant Program (2018 and 2019). The authors would like to thank Thang N. Dinh and Mo Abuhamad for their feedback on an earlier draft. 
 
 \BfPara{Authors contribution} AK and AA contributed equally; names are ordered by seniority. AK validated the generic attacks and \ours. AA developed all scripts, obtained all results, and helped analyzing them. AA drew the figures. SC and DN helped advising and providing feedback on initial ideas and helped revising the work. AA and AK wrote initial draft that was incorporated in this manuscript, written by AM. AM developed the idea of \ours. 

\balance
\bibliographystyle{IEEEtran}
\bibliography{ref.bib}

\appendix

\section{Appendix} \label{app1}
This section provides further information regarding adversarial learning algorithms utilized in this study and their categorization based on adversary's goal and knowledge.

\subsection{Adversarial Learning Algorithms}
In this study, we utilized five generic adversarial attack methods to generate visualized AEs. A brief description of these algorithms are provided in the following, and we refer more interested readers to the original works.  

\subsubsection{Fast Sign Gradient Sign Method (FGSM)} \label{sec:attacks_FGSM}
FGSM is a fast method designed to generate adversarial examples without guaranteeing similarity of the generated examples, when compared to the original samples~\cite{Goodfellow2015Explaining}. FGSM works based on a one-step gradient update, which can be expressed using:
\begin{equation}\label{eq:FGSM_Main}
    \delta= \epsilon \cdot sign \left(\nabla_{x} J_{\theta} \left( x, l \right) \right), 
\end{equation}
where $\delta$ is the perturbation parameter, $\epsilon$ is a small scalar value that limits the distortion of the image and sets the magnitude of the perturbation. Moreover, $sign(.)$ denotes the sign function, $J(.)$ is the cost function, where $x$ is the input image and $l$ is the label associated with it, and $\nabla$ computes the gradient of the cost function $J$ around the current value of $x$. Finally, the output adversarial image is by:
\begin{equation}\label{eq:FGSM_OutImage}
    \ x' =x+\delta, 
\end{equation}
where $x'$ is the adversarial example corresponding to image $x$. Controlling the distortion parameter $\epsilon$ will affect the $L_2$ norm distance between the original and the adversarial examples. A higher $\epsilon$ value corresponds to a larger $L_2$ distance. For visualization, $x'$ values are clipped to fit into [0-255] channel scale~\cite{SuVPSFS18}.

\subsubsection{Carlini \& Wagner (C\&W) Method} \label{sec:attacks_CW}
Carlini and Wagner~\cite{Carlini017} introduced three gradient based attacks by optimizing the penalty and distance metrics on $L_\infty$, $L_2$, and $L_0$ norms in the process of generating adversarial examples, as follows:
\begin{equation}\label{eq:CW_Main}
    min \, ||\delta||^2_p: \  g\left( x+\delta \right) = y' \ \& \  x+\delta \in X, 
\end{equation}
where $\delta$ is the perturbation parameter, $g(.)$ is the objective function based on the hinge loss, $y'$ is the targeted class, and $x$ is the input image. This method ensures that the added perturbation will be small and only as required to misclassify to class $y'$. A small modification can be made to launch non-targeted attacks by computing:
\begin{equation}\label{eq:CW_NonTargetted}
    min \, ||\delta||^2_p: \  g\left( x+\delta \right) \neq y \ \& \  x+\delta \in X. 
\end{equation}
In this work, we focus on the $L_2$-based C\&W attack to generate adversarial examples. The $L_2$ distance represents the required changes into the  binaries to generate the adversarial example, where a lower $L_2$ distance corresponds to a smaller perturbation in the binary to generate the example.  The perturbation $\delta$ is defined by:
\begin{equation}\label{eq:CW_Perturbation}
    \delta = \frac{1}{2}\left( tanh\left(w\right)+1\right)-x, 
\end{equation}
where $tanh(.)$ is the hyperbolic tangent function, and $w$ is an auxiliary variable optimized by: 
\begin{equation}\label{eq:CW_Wopt}
    \min\limits_{w} || \frac{1}{2}\left( tanh\left(w\right)+1\right)||_2 + c\cdot g\left(\frac{1}{2}\left(tanh\left(w\right)+1\right)\right), 
\end{equation}
where $c$ is a constant. The C\&W method minimizes the $L_p$ norm distance between the generated adversary example and the original image to increase the similarity and harden the detection process.

\subsubsection{DeepFool Method} 
DeepFool is originally designed to be untargeted iterative attack based on the $L_2$ distance metric \cite{Moosavi-Dezfooli16}. It views the neural network as a linear structure, where each class is separated with a hyperplane.  
To generate an adversarial example, DeepFool increases the distance per iteration between the input and the class associated with that input. The class is defined by:
\begin{equation}\label{eq:DEEPFOOL_Classify}
    \hat{k}(x) = arg_k \  max \  f_k(x), 
\end{equation}
where $f_k(x)$ is the output of the objective function corresponding to class $k$, which represents the class with the highest probability. Assuming that $f(x)$ is an affine classifier represented by:
\begin{equation}\label{eq:DEEPFOOL_Affine}
    f(x) = W^T x + b, 
\end{equation}
then the perturbation needed to misclassify input $x$ is described by:
\begin{equation}\label{eq:DEEPFOOL_Pertubation}
    arg_r \  min \  ||r||_2  \quad s.t. \  \exists k: w_k^T\left(x_0 + r\right) + b_k \geq w^T_{\hat{k}(x_0)} \left(x_0 + r \right) + b_{\hat{k}(x_0)}, 
\end{equation}
where $w_k$ is mapped to the $k${th} column of $W$. To generalize this method for multi-class non-linear structure, we find the next closest class to the input $x$ using the following:
\begin{equation}\label{eq:DEEPFOOL_IterativePer}
    \hat{l}\left(x_0\right) = arg\min\limits_{k \neq \hat{k}\left(x_0\right)} \frac{|f_k\left(x_0\right)-f_{\hat{k}\left(x_0\right)}\left(x_0\right)|}{||w_k - w_{\hat{k}(x_0)}||_2}. 
\end{equation}
By finding the minimum perturbation needed to misclassify the model using~\autoref{eq:DEEPFOOL_MinimalPert}, we generate the AE using $x'=x + r_*$.
\begin{equation}\label{eq:DEEPFOOL_MinimalPert}
    r_*\left(x_0\right) = \frac{|f_{\hat{l}\left(x_0\right)}\left(x_0\right)-f_{\hat{k}\left(x_0\right)}\left(x_0\right)|}{||w_{\hat{l}(x_0)} - w_{\hat{k}(x_0)}||_2^2} \left(w_{\hat{l}(x_0)}-w_{\hat{k}(x_0)}\right).
\end{equation}

\subsubsection{Momentum Iterative Method} \label{sec:attacks_MIM}
Momentum iterative method (MIM) is based on applying momentum gradient over basic FGSM to generate adversarial examples \cite{dong2018boosting}, the main goal is to maintain efficiency against black box models. The main object for this method is to find $x'$ that misclassify the model that satisfy the equation~\autoref{eq:MIM_Object} where $J$ is the loss function, and $\epsilon$ is small scalar that control the maximum distortion allowed.
\begin{equation}\label{eq:MIM_Object}
    arg \max\limits_{x'} \  J\left(x',y\right), \quad s.t. \  ||x'-x||_\infty \leq \epsilon
\end{equation}
The momentum gradient ($M_g$) then calculated using~\autoref{eq:MIM_Momentum}.
\begin{equation}\label{eq:MIM_Momentum}
    M_{g_{t+1}} = \mu M_{g_{t}} + \frac{\nabla_x J_{\theta}\left(x'_t , l\right)}{||\nabla_x J_{\theta}\left(x'_t , l\right)||}
\end{equation}
where $\mu$ is the decay factor. Initially, $x'_0$ is the original input, and $M_{g_{0}}$ is $0$. Each iteration, $x'$ will be updated according to this equation: 
\begin{equation}\label{eq:MIM_Adversary}
    x'_{t+1} = x'_t + \epsilon \cdot sign\left(M_{g_{t}}+1\right)
\end{equation}
after $n$ number of iterations, the $x'_{t+1}$ will be returned as the adversarial example for input $x$.

\subsubsection{PGD Method} \label{sec:attacks_Madry}
Madry \etal introduced an iterative projected gradient descent (PGD) attack \cite{MadryMSTV17}. The aim of PGD is to generate an adversarial sample under a minimized empirical risk with a trade of high performance cost. The original empirical risk minimization ($ERM$) for the model defined by: 
\begin{equation}\label{eq:Madry_RiskNotation}
    \mathbb{E}_{(x,y)\sim D}[L\left(x,y,\theta\right)], 
\end{equation}
where $L$ is the loss function. By modifying the definition of $ERM$ and allowing the adversary to perturb the input, $S$ is described in: 
\begin{equation}\label{eq:Madry_ERM}
    \min\limits_\theta \rho\left(\theta\right):  \rho\left(\theta\right) = \mathbb{E}_{(x,y)\sim D}[\max\limits_{\delta\in S}  L\left(x+\delta,y,\theta\right)], 
\end{equation}
where $\delta$ is the perturbation, and $\rho\left(\theta\right)$ is the objective function to minimize. As an iterative method, $x'_{t+1}$ is updated each iteration according to the previous $x'_{t}$ and the generated perturbation, after $n$ iterations, $x'_{t+1}$ will be returned as the adversarial sample. Our goal with PGD is to minimize the risk while adding perturbation in order to generate the adversarial example.

\end{document}